\documentclass[11pt]{article}
\usepackage[english,activeacute]{babel}
\usepackage{natbib}
\usepackage{comment}
\usepackage{float}
\usepackage[hidelinks]{hyperref}
\usepackage{mathrsfs}
\usepackage{enumitem}
\usepackage[font={small,it}]{caption}
\usepackage{amsmath,amsfonts,amsthm,amssymb}
\usepackage{bm,rotating,multirow,dsfont,graphicx}
\usepackage[usenames, dvipsnames]{color}
\usepackage{url}
\usepackage{multicol}
\usepackage{multirow}
\usepackage[T1]{fontenc}
\usepackage{flafter}
\usepackage{appendix}
\usepackage{subfigure}
\usepackage{xcolor}
\usepackage{soul}
\makeatletter
\def\hlinewd#1{%
	\noalign{\ifnum0=`}\fi\hrule \@height #1 %
	\futurelet\reserved@a\@xhline}
\makeatother
\newcommand{\dd}{\mathrm{d}}
\newcommand{\lp}{\left(}
\newcommand{\rp}{\right)}
\newcommand{\lb}{\left[}
\newcommand{\rb}{\right]}
\newcommand{\lc}{\left\{}
\newcommand{\rc}{\right\}}
\newcommand{\be}{\begin{eqnarray*}}
\newcommand{\ee}{\end{eqnarray*}}
\newcommand{\bet}{\begin{eqnarray}}
\newcommand{\eet}{\end{eqnarray}}
\def\spacingset#1{\renewcommand{\baselinestretch}{#1}\small\normalsize}\spacingset{1}
\def\@roman#1{\romannumeral #1}
\graphicspath{{./plots/}}

\begin{document}

\title{Bayesian Inference of Geometric Brownian Motion: An Extension with Jumps}

\date{}

\author{
    Yifei Yan, University of California, Santa Cruz, US \\
    Juan Sosa, Universidad Nacional de Colombia, Colombia \\
    Carlos Martínez, Universidad Nacional de Colombia, Colombia
}

\maketitle

\begin{abstract} 
This analysis derives the maximum likelihood estimator and applies Bayesian inference to model geometric Brownian motion, incorporating jump diffusion to account for sudden market shifts. The Bayesian approach is implemented using Markov Chain Monte Carlo simulations on S\&P 500 stock data from 2009 to 2014, providing a robust framework for analyzing stock dynamics and forecasting future trends. Exact solutions are obtained for both the standard Geometric Brownian Motion (GBM) model and the GBM model with Poisson jumps. Although both models yield reasonable results and fit the data well, the GBM with Poisson jumps exhibits superior performance, significantly enhancing model fit and capturing more complex market dynamics.
\end{abstract}

\noindent
{\it Geometric Brownian Motion, Bayesian Inference, Markov Chain Monte Carlo, Jump Diffusion, S\&P 500.}

\spacingset{1.1} 

\section{Introduction}

This paper explores the modeling of financial data using Geometric Brownian Motion (GBM) and its extension with Poisson jumps. Originally introduced by \cite{samuelson1965}, GBM has been widely adopted in financial mathematics for modeling stock prices due to its mathematical tractability and ability to represent exponential growth with stochastic fluctuations. \cite{ross} emphasized its relevance in financial modeling, particularly in option pricing. However, the assumption of continuous price paths in GBM often fails to capture the abrupt jumps and discontinuities observed in real financial markets. To address this limitation, \cite{merton} extended the GBM framework by incorporating Poisson jumps to model sudden price movements, while \cite{kou} refined this approach with a jump-diffusion model that provides greater flexibility in capturing both jump sizes and probabilities, making it more applicable to empirical data.

Here, we build upon these foundational works to derive exact solutions for both the standard GBM model and its extension with Poisson jumps. Our primary objective is to evaluate the effectiveness of these models in capturing stock price dynamics and to evaluate their predictive performance. Using Bayesian inference, we estimate model parameters and generate forecasts for the S\&P 500 stock index over a defined period of time. Bayesian methods have become increasingly important in financial econometrics, as demonstrated by \cite{geweke1999}, who highlighted their ability to incorporate prior information and account for parameter uncertainty. Within this framework, we analyze the posterior distributions of key parameters, including drift, diffusion, and jump intensity, utilizing Markov Chain Monte Carlo (MCMC) methods (e.g.,\citealt{gamerman2006markov}) for efficient sampling.

Our methodological approach begins with the derivation of the Maximum Likelihood Estimator (MLE) for the standard GBM model, using logarithmic transformations to simplify the estimation process. We then extend this framework through Bayesian hierarchical modeling \citep{gelman2013bayesian}, incorporating conjugate priors to improve computational efficiency. For GBM with jumps, we adopt the Merton's framework, modeling jumps as a Poisson process with normally distributed jump sizes on the logarithmic scale. This approach aligns with \cite{eraker2003}, who emphasized the critical role of accurately modeling jumps in financial time series. Furthermore, we employ non-informative priors, such as \(\textsf{N}(0, 100)\) for drift parameters and \(\textsf{Beta}(1,1)\) for jump probabilities, to mitigate subjective bias and ensure robustness in our analyses.

We apply these models to daily closing prices of the S\&P 500 index from 2009 to 2015, adjusting for trading days to ensure consistency in the analysis. Our results indicate that the GBM with jumps significantly outperforms the standard GBM in terms of model fit and predictive accuracy. This finding is consistent with previous studies, such as \cite{andersen2007}, which demonstrated that incorporating jumps into financial models improves their ability to capture heavy tails and skewness in asset returns. However, we also observe that the assumption of time-invariant parameters limits predictive performance. As a potential improvement, future research could extend this framework by incorporating time-dependent drift and diffusion components to better capture evolving market dynamics.

This paper is structured as follows. Section 2 derives the MLE for the standard GBM, introduces the Bayesian framework, and extends the analysis to the GBM with Poisson jumps. Section 3 applies these models to S\&P 500 data, presenting posterior summaries and predictive results. Finally, Section 4 discusses key findings, model limitations, and potential directions for future research.

\section{Modeling}

Geometric Brownian Motions (GBMs) are continuous-time stochastic processes in which the logarithm of the process follows a Brownian motion with drift \citep{ross}. These models are widely applied in mathematical finance to analyze stock price behavior and optimize option pricing. A GBM $S(t)$ is governed by the following stochastic differential equation (SDE):
\bet
\qquad\dd S(t) = \mu S(t)\dd t+\sigma S(t)\dd B(t),
     \quad t \ge 0 \,,\quad S(0) = x\,, \label{eq_sde}
\eet
where \(\mu\) denotes the drift parameter, \(\sigma\) represents the diffusion (or volatility) parameter, and \(B(t)\) is a standard Brownian motion. In practical applications, \(\mu\) and \(\sigma\) are typically unknown and must be estimated. In this study, we derive the Maximum Likelihood Estimator (MLE) for GBM and develop Bayesian hierarchical models for inference and prediction. These methods are applied to S\&P 500 stock data to assess the model’s effectiveness in capturing stock price dynamics.

\subsection{Maximum likelihood estimation}

To make inference about the model parameters, we start by solving the differential equation (\ref{eq_sde}). By assuming \( f(S_t) = \log S(t) \) and applying Ito's lemma, we obtain:
\bet 
    \dd f(S_t) &=& \frac{1}{S(t)}\dd S(t)+
        \frac{1}{2}\lp-\frac{1}{[S(t)]^2}\rp [\dd S(t)]^2\,, \notag\\
    \log S(t) &=& \log S(0)+\int_0^t \frac{\mu S_{t'}}{S_{t'}}\dd t'+\int_0^t \frac{\sigma S_{t'}}{S_{t'}}\dd B(t')-\frac{1}{2}\int_0^t\frac{[\sigma S(t')]^2 }{[S(t')]^2}
        [\dd B(t')]^2\notag\\
    &=& \log x+\int_0^t\lp\mu-\frac{\sigma^2}{2}\rp\dd t' +\int_0^t\sigma \dd B(t')\,, \notag\\
    \Rightarrow S(t) &=& x\exp\lc\lp\mu-\frac{\sigma^2}{2}\rp t
        +\sigma B(t)\rc\,. \notag
\eet

The solution suggests that \(\log S(t_0 + t) - \log S(t_0)\), for \(t_0 \geq 0\) and \(t > 0\), follows a Normal distribution. Letting \(\theta = \mu - \sigma^2 / 2\), we have that
\be \log S(t_0+t)-\log S(t_0) &=& \lp\mu-\frac{\sigma^2}{2}\rp t
        +\sigma[B(t_0+t)-B(t_0)]\\
    &=& \theta t+\sigma[B(t_0+t)-B(t_0)]\\
    &\sim& \textsf{N}\lp\theta t, \sigma^2t\rp\,,
\ee
and therefore, $B(t_1)-B(t_2)\sim \textsf{N}(0,t_1-t_2)$. Consequently, the MLE for the GBM can be obtained by applying a logarithmic transformation to the data. Let \( Y_i = \log S(t_i) \). Conditioning on the initial value, the log-likelihood functions can be expressed as
\bet 
    \log \ell(\theta,\sigma^2\mid\pmb y) &=&
    \text{c}-\frac{n}{2}\log(\sigma^2)-\sum_{i=1}^n
    \frac{\lb y_i-y_{i-1}-\theta
    (t_i-t_{i-1})\rb^2}{2\sigma^2(t_i-t_{i-1})}\,, \label{eq_ll_sde}
\eet
where $\pmb y = (y_1,\ldots,y_n)$ and $\text{c}=\sum_{i=1}^n\log(2\pi(t_i-t_{i-1}))/2$.

By taking the partial derivatives of (\ref{eq_ll_sde}) with respect to the drift and diffusion parameters, we derive the following maximum likelihood estimates for \(\theta\) and \(\sigma^2\):
$$ 
    \hat\theta = \frac{y_n-y_0}{t_n-t_0}
    \quad\text{and}\quad
    \hat\sigma^2 = \frac{1}{n}\lb\sum_{i=1}^n\frac{(y_i-y_{i-1})^2}
        {t_i-t_{i-1}}-\frac{(y_n-y_0)^2}{t_n-t_0}\rb\,.
$$
Thus, by applying the invariance property of the MLE \citep{lehmann2006theory}, the MLE of \(\mu\) can be obtained through a back-transformation, yielding \(\hat{\mu} = \hat{\theta} + \hat{\sigma}^2 / 2\).

\subsection{Bayesian modeling}\label{sec_b0_sde}

Alternatively, a Bayesian approach can be employed to estimate the two parameters. To that end, recall that:
\be 
    \log\frac{S(t_0+t)}{S(t_0)} &\sim& \textsf{N}(\theta t, \sigma^2 t)\,,
\ee
where $\theta=\mu-\sigma^2/2$. This suggests that conjugate priors can be used for modeling \(\theta\) and \(\sigma^2\). Considering again \(Y_i = \log S(t_i)\), we adopt a Normal-Inverse-Gamma prior to define the following hierarchical model:
\be y_i-y_{i-1}\mid\theta,\sigma^2
    &\sim& \textsf{N}(\theta(t_i-t_{i-1}),\sigma^2(t_i-t_{i-1}))\\
    \theta   &\sim& \textsf{N}(0,100)\\
    \sigma^2 &\sim& \textsf{IG}(2,0.001)\,.
\ee
We choose a \(\textsf{N}(0, 100)\) prior for \(\theta\) as it is non-informative, with a mean of 0 to avoid bias and a large variance accounting for large uncertainty. Similarly, we select an \(\textsf{IG}(2, 0.001)\) prior for \(\sigma^2\) to ensure a finite mean with weak prior information.

Thus, the posterior distributions of \(\theta\) and \(\sigma^2\) can be sampled using MCMC, specifically through a Gibbs sampler, following these update steps:
\begin{itemize}
    \item[1.] Sample $\theta\mid\sigma^2\sim \textsf{N}(\mu_\theta,\sigma^2_\theta)$, where
        $$
            \mu_\theta = \frac{\sum_{i=1}^n(y_i-y_{i-1})}{0.01\,{\sigma}^{2} + \sum_{i=1}^n(t_i-t_{i-1})}\,,\quad
            \sigma^2_\theta = \frac{{\sigma}^{2}}{0.01\,{\sigma}^{2} + \sum_{i=1}^n(t_i-t_{i-1})}\,.
        $$
    \item[2.] Sample $\sigma^2\mid\theta\sim \textsf{IG}(a_\sigma,b_\sigma)$, where
        $$ 
            a_\sigma = 2+\frac{n}{2}\,,\quad
            b_\sigma = 0.001+\frac{1}{2}\sum_{i=1}^n
                \frac{\left( ( y_i-y_{i-1} ) - \theta (t_i-t_{i-1}) \right)^2}{t_i-t_{i-1}}\,.
        $$
\end{itemize}

Converting \((\theta, \sigma^2)\) back to \((\mu, \sigma)\) yields the posterior distributions for the drift and diffusion parameters. Using the exact solution, we can simulate paths of \(S(t)\) from each posterior sample, enabling straightforward short-term predictions beyond the observed time horizon.

\subsection{GBMs with jumps}\label{sec_GBMj_sde}

One key assumption of GBM is that the random process follows a continuous path. However, a more general and often more realistic assumption is that the path combines both continuous and jump components. In the same spirit of \cite{merton}, we formulate a GBM with Poisson jumps, described by the following SDE:
\begin{equation}
    \frac{\dd S(t)}{S(t)} = \mu \dd t+\sigma \dd B(t) +\dd\lb\sum_{j=1}^{N_t}(V_j-1)\rb,
    \quad t\ge0\,, \quad S(0)=x \,,\label{eq_sde_2}
\end{equation} 
where \(N_t\) is a point process (e.g., \citealt{snyder2012random}) representing jumps, where \(N_t - N_s \sim \textsf{Poisson}(\lambda(t-s))\), and \(V_j\) are independent and identically distributed (iid) jump sizes from a specified distribution. The analytical solution of \eqref{eq_sde_2} is available in closed form. Here, we present the result for the simplest SDE involving jump processes, accompanied by a sketch of the proof.

As in \cite{privault}, consider the following differential equation:
\be S(t) = \eta S(t^-) \, \dd N_t\,,
    \quad\eta\in\mathbb R\,, 
    \quad S(0)=x\,,
    \quad \dd N_t = N(t)-N(t^-)=1\,,
\ee
meaning that there is exactly one jump at time $t$. Thus, the above equation simplifies to $\dd S(t) = S(t)-S(t^-)=\eta S(t^-)$. After rearranging terms and completing the proof by induction, we obtain $S(t) = (1+\eta)S(t^-)$, which implies that $S(t) = x(1+\eta)^{N_t}$. Generalizing this to a non-constant \(\eta\) provides the solution to (\ref{eq_sde_2}) as follows:
$$ 
    S(t) = x\exp\lc\lp\mu-\frac{\sigma^2}{2}\rp t + \sigma B(t)\rc\prod_{j=1}^{N_t}V_j = x\exp\lc\lp\mu-\frac{\sigma^2}{2}\rp t + \sigma B(t)+\sum_{j=1}^{N_t}Z_j\rc\,,
$$
where $Z_j=\log V_j$.

\cite{kou} shows that when \(\Delta t\) is sufficiently small, the jump component in the exponent can be approximated in distribution by a mixture of \(Z_t\)'s, weighted by Bernoulli variables \(J_t\) with success probability \(\lambda \Delta t\), where \(\lambda\) represents the jump rate. Taking the logarithm of the data and assuming a common \(\textsf{N}(\mu_z, \sigma^2_z)\) distribution for the \(Z_t\)'s, we derive the following transition distribution for equally spaced data:
\be y_i-y_{i-1} &=& \log S(t_i)-\log S(t_{i-1}) \\
    &\approx& \lp\mu-\frac{\sigma^2}{2}\rp (t_i-t_{i-1}) +\sigma \left(B(t_i)-B(t_{i-1})\right) + \sum_{t=t_{i-1}}^{t_i}J_t Z_t \\
    &\sim& \textsf{N}(\theta\Delta t, \sigma^2\Delta t) + \textsf{Ber}(\lambda^\ast)\cdot \textsf{N}(\mu_z,\sigma^2_z)\,,
\ee
where $\Delta t=t_i-t_{i-1}$ and $\lambda^\ast=\lambda\Delta t$. 

This result is highly practical, as it enables efficient sampling of \(\theta\), \(\sigma\), and \(\lambda\) using MCMC. Reapplying a \(\textsf{N}(0, 100)\) prior for \(\mu_z\) and an \(\textsf{IG}(2, 0.001)\) prior for \(\sigma_z^2\) ensures conjugate updates for the hyperparameters. Since \(\lambda^\ast\) represents the parameter of a Bernoulli random variable, a \(\textsf{Beta}(1, 1)\) prior is a natural non-informative choice. With conjugate priors assigned to all model parameters, the full conditional distributions can be derived analytically. For the sake of brevity, we omit the details of the MCMC algorithm, as it closely resembles the one outlined previously.

\section{Illustration}

\begin{figure}[!htb]\vspace{-3em}
    \centering
    \includegraphics[scale=0.75]{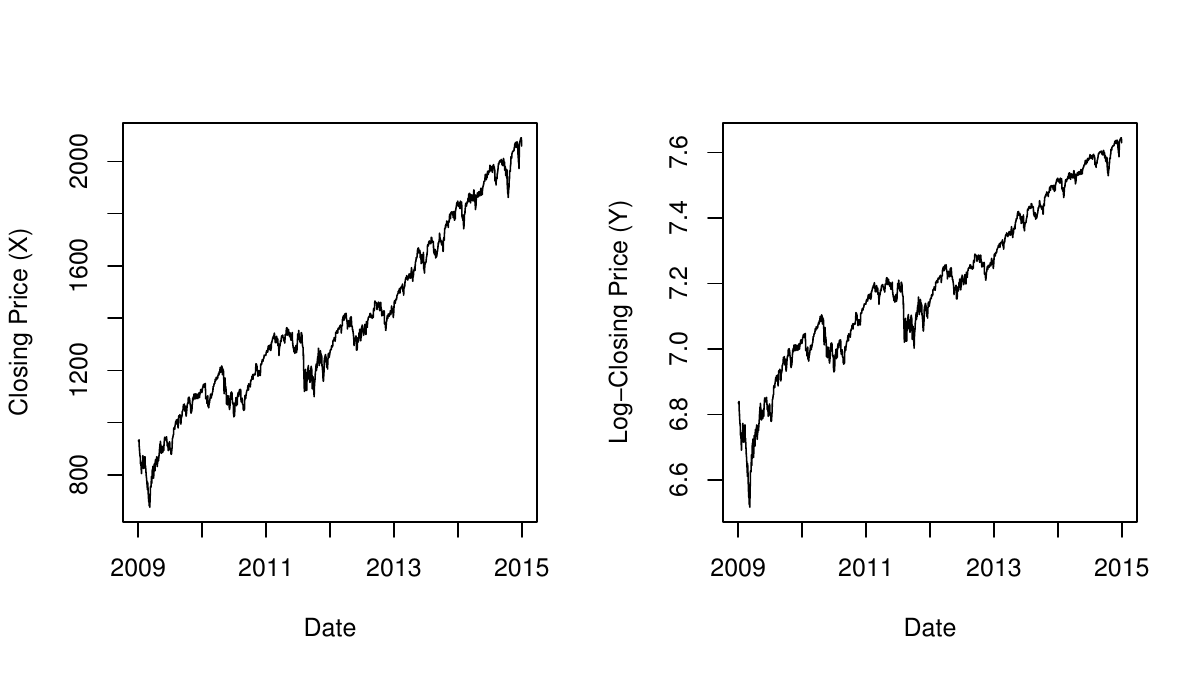}\vspace{-1em}
    \caption{Raw and log-transformed closing price of S\&P 500 over 2009-2015.}\label{fig_raw_sde}
\end{figure}

We use five years of consecutive daily closing price data for the S\&P 500, spanning from January 2, 2009 (the first trading day of 2009), to December 31, 2014. The S\&P 500 is a prominent American stock market index comprising the market capitalizations of 500 large companies listed on the New York Stock Exchange (NYSE) or the National Association of Securities Dealers Automated Quotations (NASDAQ). Figure \ref{fig_raw_sde} plots of the raw data (\(X\)) and its log-transformed counterpart (\(Y\)).

\begin{table}[!t]
    \centering
    \begin{tabular}{l c c c c c c}
    \hline
    & &\multicolumn{5}{c}{\textit{Posterior Distribution}}\\
    \cline{3-7}
    \textit{Parameter} & \textit{MLE} &
        \textit{Mean} & \textit{SD} &
        \textit{2.5\%} & \textit{50\%} & \textit{97.5\%}\\
    \hline
    Drift: $\mu$ & 0.149 & 0.150 & 0.075 & 0.003 & 0.150 & 0.298\\
    Difussion: $\sigma$ & 0.183 & 0.183 & 0.003 & 0.177 & 0.183 & 0.190\\
    \hline
    \end{tabular}
    \caption{MLE and the posterior distribution of $\mu$ and $\sigma$.}\label{tab_sde_1}
\end{table}

\begin{figure}[!htb]
    \centering
    \includegraphics[scale=0.67]{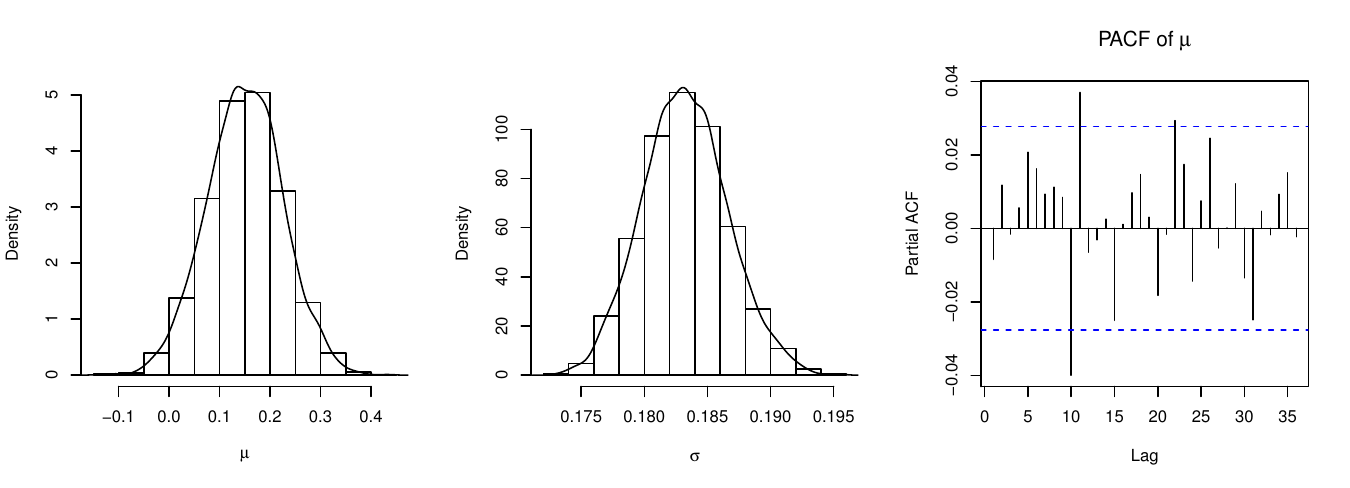}\vspace{-1em}
    \caption{Histogram and posterior density of $\mu$ and $\sigma$, and PACF of $\mu$.}\label{fig_sde_2}
\end{figure}

We select a year as the unit of time and scale the days by a factor of \(1/252\), reflecting the approximately 252 trading days in a calendar year. Using 5,000 simulations following a burn-in period of 1,000 iterations, we fit a GBM to the S\&P 500 data, as described in Section \ref{sec_b0_sde}. The summary statistics for the 5,000 samples of the drift and diffusion parameters are presented in Table \ref{tab_sde_1}. The drift parameter is positive, with its 95\% credible interval excluding 0, suggesting an upward trend in the stock price. The diffusion parameter is highly stable, with a tightly concentrated posterior and a coefficient of variation as low as 0.018. This stability suggests that the stock index's volatility does not vary significantly over time, which is reasonable given that the S\&P 500 is a composite index representing 500 stocks. The histograms and partial autocorrelation function (PACF) of the posterior samples, shown in Figure \ref{fig_sde_2}, indicate that the algorithm performs effectively.

\begin{figure}[!htb]
    \centering
    \includegraphics[scale=0.77]{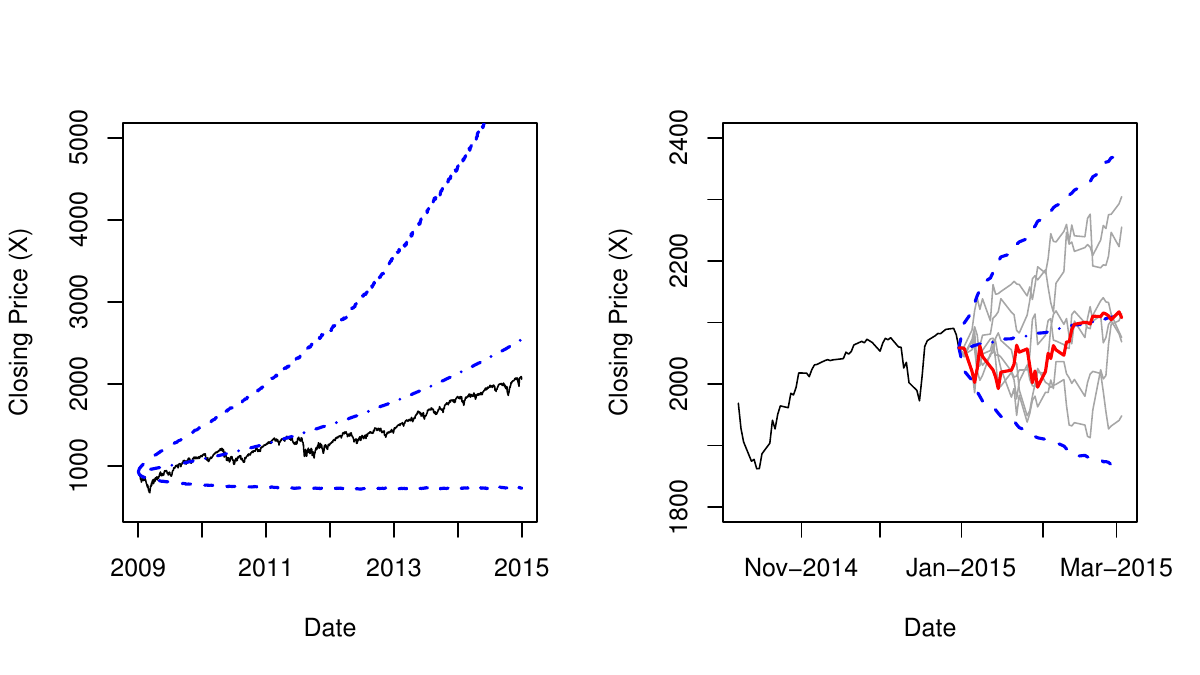}\vspace{-1em}
    \caption{Posterior mean and 90\% credible interval of fitted realizations for 2009 through 2014 under the GBM, with true values highlighted in red.}\label{fig_sde_3}
\end{figure}

Figure \ref{fig_sde_3} shows the posterior mean and 90\% credible interval of the fitted realizations of 2009 through 2014. Although the fitted process missed the first big dip at the beginning of the time series, it is consistent with the general behavior of the truth. Using data from 2009 to 2014, we predict the closing price of S\&P 500 in the first two months of 2015 and compare it with the true value. Again, the credible interval captures the actual stock behavior quite well.
Also, notice that the 90\% credible band is relatively wide for both the fitted and predictive processes. This can be attributed to the volatility (mean 0.18) being larger compared to the drift (mean 0.15). Additionally, the predicted paths appear more volatile than the actual process. This discrepancy is likely due to the original process occasionally exhibiting sharp dips, which are not captured by the GBM. Since the variability of these dips is absorbed into the standard diffusion estimation, it is not surprising that the predictions exhibit seemingly inflated volatility compared to the observed data.

\begin{table}[!b]
    \centering
    \begin{tabular}{l c c c c c c c}
    \hline
    & \textit{GBM} &&\multicolumn{5}{c}{\textit{GBM with Jumps}}\\
    \cline{2-2}\cline{4-8}
    \textit{Parameter} & \textit{Mean} &&
        \textit{Mean} & \textit{SD} &
        \textit{2.5\%} & \textit{50\%} & \textit{97.5\%}\\
    \hline
    \underline{GBM component}\\
    ~~Drift: $\mu$ & 0.150 && 0.349 & 0.062 & 0.229 & 0.348 & 0.473\\
    ~~Difussion: $\sigma$ & 0.183 && 0.089 & 0.005 & 0.078 & 0.089 & 0.099\\
    \underline{Jump component}\\
    ~~Mean: $\mu_Z$ & -- && -0.002 & 0.001 & -0.004 & -0.002 & 0.000\\
    ~~SD: $\sigma_Z$ & -- && 0.017 & 0.001 & 0.015 & 0.017 & 0.019\\
    ~~Probability: $\lambda^\ast$ & -- && 0.360 & 0.038 & 0.288 & 0.360 & 0.438\\
    \hline
    \end{tabular}
    \caption{Posterior distributions of model parameters for both the standard GBM and the GBM with jumps. The 2.5\%, 50\%, and 97.5\% values represent the corresponding quantiles of the posterior distribution.}\label{tab_sde_2}
\end{table}

Now, we model the data using the GBM with jumps framework described in Section \ref{sec_GBMj_sde}. Table \ref{tab_sde_2} presents the posterior summary statistics under this model, compared to the earlier results from the simple GBM without jumps. Accounting for potential jumps in the process significantly reduces the diffusion estimate of the GBM component to 0.089. Similarly, the diffusion-to-drift ratio drops from 1.23 to 0.26, indicating much lower variability in the GBM component. The posterior mean of the jump probability per trading day is estimated at 0.36, a substantial value. This highlights the importance of including the jump component, as omitting it would overlook a critical aspect of the process's variability.

\begin{figure}[!htb]\vspace{-3em}
    \centering
    \includegraphics[scale=0.77]{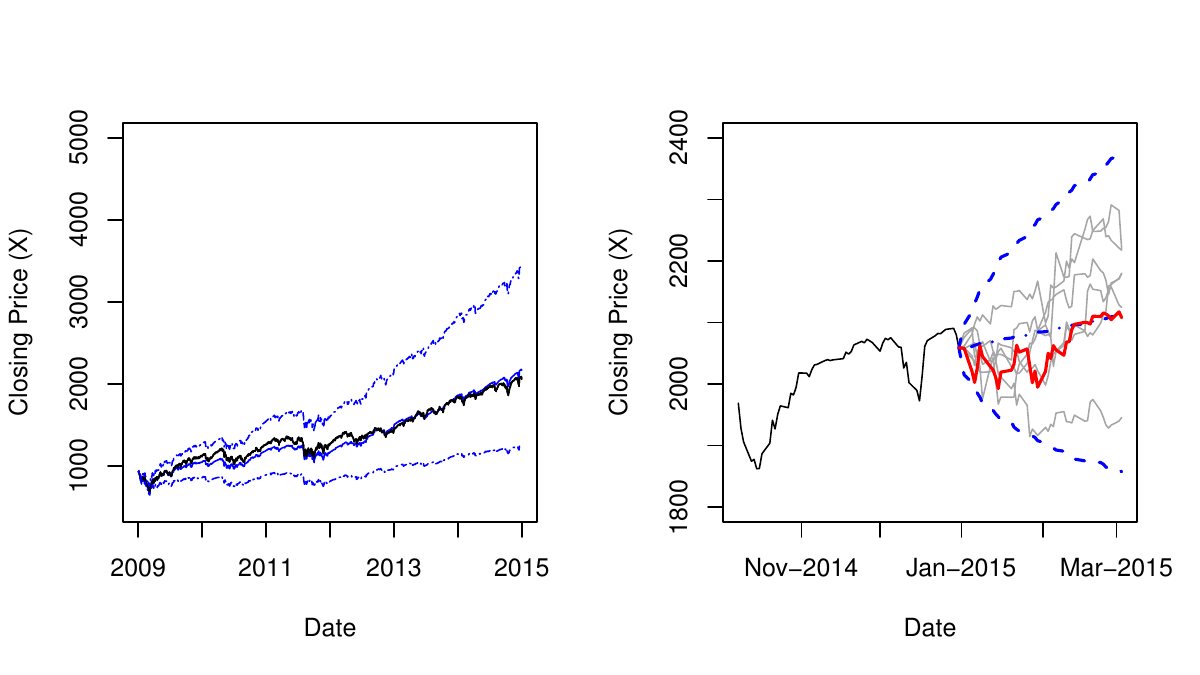}\vspace{-1em}
    \caption{The posterior mean and the 90\% credible interval under the GBM with jumps. True values indicated in red.}\label{fig_sde_4}
\end{figure}

Finally, we evaluate the fitted process from 2009 to 2014 and the posterior predictive paths for the first two months of 2015, as shown in Figure \ref{fig_sde_4}. Visually, the GBM with jumps clearly provides a superior fit compared to the simple GBM. The new model accurately captures key features of the data, including the sharp dip in early 2009 at the start of the time series. Moreover, it produces a much narrower credible band for the fitted paths, reflecting reduced uncertainty compared to the simple GBM. For predictions in 2015, while the prediction bands are similar to those from the previous analysis, the variability diminishes in segments without obvious jumps. Additionally, the predicted paths under the GBM with jumps align more closely with the observed data, showcasing improved predictive performance.

Both models provided reasonable results, with the GBM with jumps offering a significantly better fit. While both models produced similar short-term predictions, incorporating time-dependent parameters could enhance accuracy. The assumption of a Normal distribution for jump sizes could be replaced by alternative distributions, like a t distribution or a double-exponential, for better tail behavior modeling. The model can also be adapted for non-equidistant data, although this requires a Metropolis-Hastings step for sampling the jump rate \(\lambda\). Future research should explore the impact of alternative prior specifications on posterior estimates, particularly for jump parameters, and incorporate informative priors based on market data to improve precision. Extending the model to multivariate processes and integrating machine learning techniques, such as Gaussian processes or neural networks, could improve predictive capabilities. Lastly, improving computational efficiency for high-frequency data and complex models is crucial, and efficient sampling algorithms or parallel computing could significantly reduce computation time while maintaining robust posterior inference.

\section{Discussion}

In this analysis, we derived exact solutions for both the standard GBM and the GBM with Poisson jumps, applied Bayesian methods to estimate posterior distributions of the model parameters, and generated predictions for the S\&P 500 stock prices from 2009 to 2015. While both models produced reasonable results and demonstrated a good fit to the data, the GBM with jumps exhibited a significantly improved fit, capturing more complex market dynamics.

An important avenue for future research is investigating the impact of alternative prior specifications on posterior estimates, particularly for the jump intensity and jump size parameters. While this study employed non-informative priors for simplicity, incorporating informative priors derived from historical market data could enhance the precision of parameter estimates in specific contexts. Additionally, extending the model to accommodate multivariate processes—such as the simultaneous modeling of multiple stock indices or asset classes—could provide deeper insights into interdependencies and co-movement patterns in financial markets. Another promising direction involves integrating machine learning techniques, such as Gaussian processes or neural networks, to improve the predictive performance of the Bayesian framework. Furthermore, computational efficiency remains a critical challenge, particularly for high-frequency data or models with complex hierarchical structures. Advancing more efficient sampling algorithms or leveraging parallel computing could significantly reduce computation time while preserving the robustness of posterior inference.

\section*{Statements and Declarations}

The authors declare that they have no known competing financial interests or personal relationships that could have influenced the work reported in this article.  

During the preparation of this work, the authors used ChatGPT-4-turbo to enhance language clarity and readability. Following the use of this tool, the authors carefully reviewed and edited the content as needed and take full responsibility for the final version of the publication.

\bibliography{references.bib}

\begin{thebibliography}{}

\bibitem[Andersen et~al., 2007]{andersen2007}
Andersen, T.~G., Bollerslev, T., and Diebold, F.~X. (2007).
\newblock Practical volatility and correlation modeling for financial market
  risk management.
\newblock {\em Handbook of Financial Time Series}, pages 777--810.

\bibitem[Eraker et~al., 2003]{eraker2003}
Eraker, B., Johannes, M., and Polson, N. (2003).
\newblock The impact of jumps in volatility and returns.
\newblock {\em Journal of Finance}, 58(3):1269--1300.

\bibitem[Gamerman and Lopes, 2006]{gamerman2006markov}
Gamerman, D. and Lopes, H.~F. (2006).
\newblock {\em Markov chain Monte Carlo: stochastic simulation for Bayesian
  inference}.
\newblock Chapman and Hall/CRC.

\bibitem[Gelman et~al., 2013]{gelman2013bayesian}
Gelman, A., Carlin, J.~B., Stern, H.~S., Dunson, D.~B., Vehtari, A., and Rubin,
  D.~B. (2013).
\newblock {\em Bayesian Data Analysis}.
\newblock Chapman and Hall/CRC, Boca Raton, FL, 3rd edition.

\bibitem[Geweke, 1999]{geweke1999}
Geweke, J. (1999).
\newblock Using simulation methods for bayesian econometric models: Inference,
  development, and communication.
\newblock {\em Econometric Reviews}, 18(1):1--73.

\bibitem[Kou, 2002]{kou}
Kou, S.~G. (2002).
\newblock A jump-diffusion model for option pricing.
\newblock {\em Management science}, 48(8):1086--1101.

\bibitem[Lehmann and Casella, 2006]{lehmann2006theory}
Lehmann, E.~L. and Casella, G. (2006).
\newblock {\em Theory of point estimation}.
\newblock Springer Science \& Business Media.

\bibitem[Merton, 1976]{merton}
Merton, R.~C. (1976).
\newblock Option pricing when underlying stock returns are discontinuous.
\newblock {\em Journal of financial economics}, 3(1-2):125--144.

\bibitem[Privault, 2013]{privault}
Privault, N. (2013).
\newblock Notes on stochastic finance.
\newblock {\em Retrieved from}.

\bibitem[Ross, 2014]{ross}
Ross, S.~M. (2014).
\newblock {\em Introduction to probability models}.
\newblock Academic press.

\bibitem[Samuelson, 1965]{samuelson1965}
Samuelson, P.~A. (1965).
\newblock Proof that properly anticipated prices fluctuate randomly.
\newblock {\em Industrial management review}, 6(2):41--49.

\bibitem[Snyder and Miller, 2012]{snyder2012random}
Snyder, D.~L. and Miller, M.~I. (2012).
\newblock {\em Random point processes in time and space}.
\newblock Springer Science \& Business Media.

\end{thebibliography}
\bibliographystyle{apalike}

\end{document}